\documentclass[doublecol]{epl2}

\usepackage{amsmath,amssymb,mathrsfs}
\usepackage{graphicx}
\usepackage{latexsym}

\allowdisplaybreaks[4]

\begin{document}
\title{Non-equilibrium current and relaxation dynamics of a charge-fluctuating quantum dot}
\author{C. Karrasch,\inst{1} S. Andergassen,\inst{1} M. Pletyukhov,\inst{1} D. Schuricht,\inst{1} L. Borda,\inst{2} V.Meden,\inst{1} \and H. Schoeller\inst{1}}

\shortauthor{C.~Karrasch \etal}

\institute{
\inst{1} Institut f\"ur Theoretische Physik A  and JARA-Fundamentals of Future Information Technology, RWTH Aachen University, D-52056 Aachen, Germany \\
\inst{2} Physikalisches Institut, Universit\"at Bonn, D-53115 Bonn, Germany}

\abstract{
  We study the steady-state current in a minimal model for a quantum dot
  dominated by charge fluctuations and analytically describe the time
  evolution into this state. The current is driven by a finite bias voltage $V$
  across the dot, and two different renormalization 
  group methods are used to treat small to intermediate local Coulomb interactions. The corresponding flow equations can be solved analytically which
  allows to identify all microscopic cutoff scales. Exploring the entire parameter space  
  we find rich non-equilibrium physics  which cannot be understood by simply 
  considering the bias voltage as an infrared cutoff. For the experimentally relevant case of left-right 
  asymmetric couplings, the current generically
  shows a power-law suppression for large $V$. 
  The relaxation dynamics towards the steady state features characteristic 
  oscillations as well as an interplay of exponential and power-law decay.}
\maketitle


\section{Introduction}
\begin{figure}[t]
\centering
\includegraphics[width=0.8\linewidth,clip]{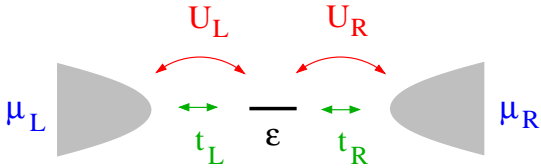}
\caption{The interacting resonant level model discussed in this work.}
\label{figmodel}
\end{figure}

Recent progress in the ability to engineer
nanostructured devices has opened new possibilities for studying the
finite-bias transport characteristics of such systems. As the electrons
occupying the nanostructure are spatially confined, local Coulomb correlations
strongly affect the physics, and understanding non-equilibrium phenomena in
systems with local two-particle interactions is therefore of fundamental importance. In
an attempt to investigate simplified cases first, one can distinguish between
situations in which either charge or spin fluctuations dominate. The latter case is described 
by the Kondo model, and progress in understanding its non-equilibrium physics was made 
recently (for a review see Ref.~\cite{S}). We here consider the other situation and study 
a minimal model for a quantum dot dominated by charge
fluctuations---the interacting resonant level model (IRLM). It
describes a spinless localized level at energy $\epsilon$ coupled to two leads
by electron hoppings $t_\alpha$ and local Coulomb repulsions $u_\alpha$ (see Fig.~\ref{figmodel}). The
lead electrons are assumed to be (effectively) non-interacting and held at two
different chemical potentials $\mu_\alpha=\pm V/2$, with $\alpha=L,R$ denoting
the left and right lead and $V$ being the bias voltage.  

The steady-state current $I$ of
the IRLM was studied intensively during the last few years using
various techniques including the scattering Bethe Ansatz~\cite{B},
perturbative and numerical renormalization group (NRG) methods ~\cite{BSZ},
the Hershfield $Y$-operator~\cite{D}, the time-dependent density matrix
renormalization group (tDMRG) method~\cite{BSS} as well as sophisticated
field theory approaches~\cite{BS,BSS}.
These studies were mostly performed at the non-generic point of 
particle-hole and left-right symmetry, which can hardly be realized in experiments. 
It was concluded that at sufficiently large $V$, $I$ decreases as a power law. Similar 
power laws were found in equilibrium and it was 
suggested that $V$ is just another infrared energy cutoff (in addition 
to, e.g., $t_\alpha$ or temperature), leading to
the speculation that the IRLM does not contain any interesting 
non-equilibrium physics \cite{BSZ}. 
Exploring the entire parameter space we show analytically that this 
conclusion is too restrictive. We uncover rich non-equilibrium physics beyond 
the situation where the voltage $V$ acts as a simple low-energy cutoff associated with a power-law behavior of the current. 
However, in the limit of strong left-right asymmetry, which can be easily realized 
experimentally, we find generic power-law scaling of $I$ for large $V$, in particular also away from particle-hole symmetry.

In addition, we provide an analytic description of the relaxation dynamics
of the system into the steady state after switching on the
level-lead coupling at time $t=0$. Two different relaxation rates control the exponential decay
which is 
accompanied by oscillatory behavior with a voltage-dependent frequency and
power-law decay with an exponent depending on $u_\alpha$.
Describing the time evolution of a
locally correlated electron system is as challenging as
understanding the non-equilibrium steady state current. Various numerical
techniques like time-dependent NRG~\cite{TD-NRG} and tDMRG~\cite{TD-DMRG}, an
iterative path-integral method~\cite{W}, and a non-equilibrium Monte Carlo
approach~\cite{Sch} were developed. Certain exactly solvable models were discussed~\cite{E}, 
and a perturbative renormalization group (RG) method~\cite{KS,PSS} as well as a flow equation
approach~\cite{KT} were applied. However, these studies do not cover 
charge-fluctuating, correlated quantum dots.

\begin{figure}[t]
  \centering
  \includegraphics[width=0.95\linewidth,clip]{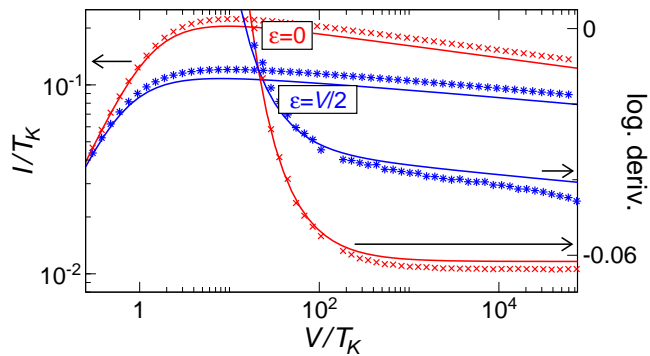}
  \caption{(Color online) $I(V)$ for the symmetric model with 
    $U_L=U_R=0.1/\pi$ obtained from the numerical solution of the full RTRG-FS
    (lines) and FRG (symbols) equations; crosses: $\epsilon=0$; stars: $\epsilon=V/2$.} 
  \label{fig:fig1}
\end{figure}

In this Letter, we use two RG methods to investigate
the IRLM. While both are bound to the case of weak Coulomb
interactions, they are complementary in other aspects. Within the 
functional RG (FRG), which was recently extended to non-equilibrium~\cite{FRGnonequi}, 
the steady state can be studied for arbitrary system parameters. In particular, this 
allows for a comparison to highly accurate tDMRG data obtained for hoppings which are too
large to be deep in the scaling limit~\cite{BSS}, the latter being realized for large 
band width and small $t_\alpha$. For small interactions, we find
excellent agreement (see Fig.~\ref{fig:fig2}(a)). In the scaling limit, the
FRG results and the ones obtained by the real-time renormalization group in
frequency space (RTRG-FS)~\cite{S} coincide (see Figs.~\ref{fig:fig1} and
\ref{fig:fig3}).  The latter method was earlier applied to systems dominated
by spin fluctuations~\cite{RTRGKondo}.
In contrast to the FRG, RTRG-FS can only be used in the scaling limit, but on the 
other hand allows for an analytical description not only of the steady state but also of 
the relaxation dynamics.  The combined use of both RG approaches leads to a reliable and 
comprehensive picture of the non-equilibrium physics under consideration. In particular, 
we identify the various microscopic cutoff scales, which is essential for 
the precise determination of the scaling behavior of observables.

\section{Model and RG equations}The Hamiltonian of the IRLM (see Fig.~\ref{figmodel}) is given by
$H=H_{l}+H_d+H_{c}$, where $H_{l}=\sum_{k\alpha}(\epsilon_k+\mu_\alpha)
a_{k\alpha}^\dagger a_{k\alpha}$ describes two semi-infinite fermionic leads
which are held at $\mu_{L/R}=\pm V/2$, respectively. Standard second quantized
notation is used, and the energies $\epsilon_{k\alpha}$ are restricted to a
finite band of width $B$.  In the scaling limit, the details of the frequency
dependence of the lead local density of states $\rho_{\alpha}$ do
not play any role as long as it is sufficiently regular for energies of the order of
$V$ and smaller. When comparing to tDMRG data \cite{BSS}, we employ the
semi-circular $\rho_\alpha(\omega)$ associated with simple tight-binding chains 
(which are used in tDMRG). The dot Hamiltonian reads $H_d=\epsilon \hat n $ 
with $\hat n =c^\dagger c$, and this single fermionic level is coupled to the leads via 
$H_c=\sum_{k\alpha}t_\alpha (a_{k\alpha}^\dagger
c+{\rm H.c.}) + (\hat{n}-\frac{1}{2})\sum_{kk'\alpha}u_\alpha
{:\!a_{k\alpha}^\dagger a_{k'\alpha}\!:}$, where $:\ldots:$ denotes
normal-ordering.  We stress that in contrast to other studies, the coupling to the
leads is allowed to be asymmetric, which is the situation generically expected
in experiments.  Furthermore, we do not only focus on the particle-hole symmetric
point $\epsilon=0$.

Within both RG approaches, coupled differential equations for the flow of the
effective system parameters as a function of an infrared cutoff $\Lambda$ holding up to leading order in $u_{\alpha}$ can 
be derived.
Aiming at an analytic discussion, it is instructive to  
consider {\it simplified} flow equations for the renormalized steady-state 
rates $\Gamma_{\alpha}$ whose bare values are given by $\Gamma_\alpha^0=2\pi 
\rho_\alpha t_\alpha^2$. In the scaling limit, both RG approaches give the 
same functional form
\begin{equation}
\frac{d \Gamma_{\alpha}}{d \Lambda}=
-2U_{\alpha}\Gamma_{\alpha}\frac{\Lambda+\Gamma/2}
{(\mu_{\alpha}-\epsilon)^2+(\Lambda+\Gamma/2)^2},
\label{eq:RGequation}
\end{equation}
with $U_\alpha=\rho_\alpha u_\alpha$ being the dimensionless interaction, 
and $\Gamma=\sum_{\alpha}\Gamma_{\alpha}$. The renormalization
of the level position $\epsilon$ is small and will be neglected. 
The RG flow
\eqref{eq:RGequation} is cut off at the scale
$\Lambda_c=\max\{|\mu_{\alpha}-\epsilon|,\Gamma/2 \}$, and an approximate
solution for $\Gamma_{\alpha}$ is given by
\begin{equation}
\label{eq:flow}
\Gamma_{\alpha}=
\Gamma_{\alpha}^0\left(\frac{\Lambda_0}{\Lambda_c}\right)^{2U_{\alpha}},
\end{equation}
where $\Lambda_0 \sim B$ denotes the initial cutoff.  At large voltages
$V\gg\Gamma$, we distinguish between the {\it off-resonance} $|V- 2 \epsilon|>\Gamma$
and the {\it on-resonance} $V = 2 \epsilon$ situation 
(peak in conductance; see Fig.~\ref{fig:fig3}).  In the latter
case, the relevant energy scales cutting off the flow are $\Gamma/2$ for
$\Gamma_L$ and $V$ for $\Gamma_R$.

Similiar to the Kondo model the cutoff parameter
is the maximum of the distance to the resonance 
and the corresponding decay rate, i.e. in our
case $\textnormal{max}(|\epsilon \pm V/2|,\Gamma)$. There
is, however, an important difference.
For the Kondo model, even at resonance $V=h$ (the latter being the magnetic
field), there is a weak-coupling expansion parameter, namely
the dimensionless exchange coupling cut off at $\textnormal{max}(V,h)$.
For the IRLM, at $\epsilon = \pm V/2$, the tunneling is not 
a weak-coupling expansion parameter since $\Gamma$ is not 
dimensionless. This fact constitutes an essential difference between
the description of resonance phenomena in models with
charge and spin fluctuations.

\begin{figure}[t]
  \centering
  \includegraphics[width=0.9\linewidth,clip]{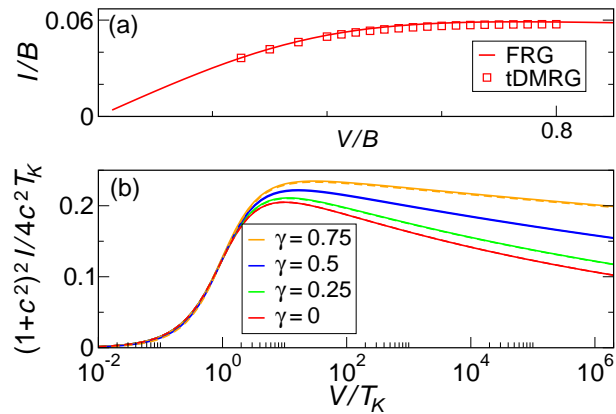}
  \caption{(Color online) a) Comparison of FRG results (lines) and tDMRG 
    data~\cite{BSS} (symbols) of $I(V)$ for $u_L=u_R=0.3B/4$ and $t_L=t_R=0.5B/4$ at 
    $\epsilon=0$. b) RTRG-FS results for $I(V)$ for $U_{L/R}=(1\pm \gamma)\,0.1/\pi$, 
    and $\epsilon=0$; analytic result \eqref{eq:flow} inserted in 
    \eqref{eq:curr} (dashed lines) mostly hidden by 
    solution of the full flow equations (solid lines); $\gamma=0.75$, $0.5$, $0.25$, $0$ (corresponding to $c^2=21.4$, $7.9$, $2.8$, $1$) from top to bottom.}
  \label{fig:fig2}
\end{figure}

The {\it full} FRG and RTRG-FS flow equations are 
presented in Refs.~\cite{KPBM} and \cite{RTlang}, respectively and can 
easily be solved on a computer. If not mentioned otherwise, the results shown in 
the Figures were obtained in this way (for a comparison to the analytic solution 
of the simplified equations see Fig.~\ref{fig:fig2}(b)). 

In the scaling limit where $\Lambda_0 \to \infty$ and $\Gamma_{\alpha}^0 \to
0$, the dependence on bare parameters vanishes, and all quantities can be
expressed in terms of the invariant scale $T_K=\sum_{\alpha}T_K^{\alpha}$,
with $T_K^{\alpha} = \Gamma_{\alpha}^0 (2\Lambda_0 / T_K)^{2 U_{\alpha}}$, and 
the asymmetry parameter $c^2 = T_K^L/T_K^R$.
Thus, at $V=0$
\begin{equation}
\Gamma=T_K\left[c\left(\frac{T_K}{\Gamma}\right)^{2U_L}
        +\frac{1}{c}\left(\frac{T_K}{\Gamma}\right)^{2U_R}\right]\frac{c}{1+c^2},
\label{eq:Gammaeq}
\end{equation}
which has the solution $\Gamma = T_K$ in the symmetric case ($U_L = U_R$,
$c=1$). The corresponding equation for $\Gamma$ at finite $V$ in the off-
(on-) resonance situation is obtained by replacing $\Gamma$'s on the
right-hand side of \eqref{eq:Gammaeq} by $V - 2 \epsilon$ and $V +2 \epsilon$
(by $\Gamma$ and $2 V$).  As a result, the rates $\Gamma_{\alpha}$ are
generically characterized by power laws with interaction-dependent exponents.


\section{Steady-state quantities}
\begin{figure}[t]
  \centering
  \includegraphics[width=0.95\linewidth,clip]{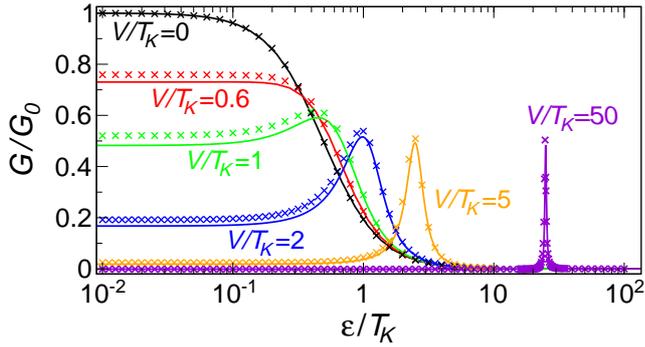}
  \caption{(Color online)  Conductance $G(\epsilon)=dI/dV$ (lines:
    RTRG-FS, symbols: FRG) in the symmetric model with $U_L=U_R=0.1/\pi$.}
  \label{fig:fig3}
\end{figure}

The dot occupation in the stationary state
reads
\begin{equation}
  \langle\hat n\rangle =\frac{1}{2}+\frac{1}{\pi}
  \left[\frac{\Gamma_L}{\Gamma}\,
    \text{arctan}\frac{V -2 \epsilon}{\Gamma}-
    \frac{\Gamma_R}{\Gamma}\,
    \text{arctan}\frac{V +2 \epsilon}{\Gamma}\right],
  \label{eq:n}
\end{equation}
and the static susceptibility is defined as $\chi =-\frac{\partial \langle\hat
  n \rangle}{\partial\epsilon} |_{\epsilon=0}$. In the symmetric case and at
$V=0$, one obtains $\chi^{\rm{sym}}_{V=0}=2/(\pi \Gamma)$~\cite{chinrg}, which
can be used to define the physical scale $T_K = \frac{2}{\pi}
(\chi^{\rm{sym}}_{V=0})^{-1}$ even away from the scaling limit. The stationary current can directly be
computed from the rates $\Gamma_\alpha$:
\begin{equation}
\label{eq:curr}
I=\frac{1}{\pi}\frac{\Gamma_L\Gamma_R}{\Gamma}
        \left[\text{arctan}\frac{V-2 \epsilon}{\Gamma}+
        \text{arctan}\frac{V+ 2 \epsilon}{\Gamma}\right].
\end{equation}
For $V\gg\Gamma$ and {\it off resonance,} this expression simplifies to $I
\approx \Gamma_L\Gamma_R/\Gamma$ and thus
\begin{equation}
  I(V) \approx 
  T_K\frac{\left(\frac{T_K}{|V- 2 \epsilon|}\right)^{2U_L}
    \left(\frac{T_K}{|V +2 \epsilon|}\right)^{2U_R}}
        {c\left(\frac{T_K}{|V - 2 \epsilon|}\right)^{2U_L}
        +\frac{1}{c}\left(\frac{T_K}{|V + 2 \epsilon|}\right)^{2U_R}}\frac{c}{1+c^2}.
      \label{eq:IVoff}
\end{equation}
Whereas for $U_L=U_R=U$ and $V\gg\epsilon$ (e.g., at the particle-hole symmetric point $\epsilon=0$) 
the current is always governed by a power law
$I(V)\propto V^{-2U}$ in agreement with earlier
studies~\cite{D,BSS}, this does not hold in general for asymmetric Coulomb
interactions generically realized in experiments. In this case the two 
terms in the denominator of (\ref{eq:IVoff})
are typically of the same order of magnitude. Only if in addition 
to $U_L\neq U_R$ the asymmetry in the bare rates is large 
($c \ll 1$ or $c \gg 1$), the power-law behavior of $I(V)$ is 
recovered (with exponents $2U_L$ or $2U_R$, respectively). In the 
{\it on-resonance} case $\epsilon=V/2$ where the conductance $G=dI/dV$ 
has a maximum (see Fig.~\ref{fig:fig3}), 
the current is given by
\begin{equation}
  I(V) \approx \frac{\Gamma_L \Gamma_R}{2 \Gamma} 
= \frac{T_K}{2}\frac{\left(\frac{T_K}{\Gamma}\right)^{2U_L}
    \left(\frac{T_K}{2V}\right)^{2U_R}}
  {c\left(\frac{T_K}{\Gamma}\right)^{2U_L}
    +\frac{1}{c}\left(\frac{T_K}{2V}\right)^{2U_R}}\frac{c}{1+c^2}.
\label{eq:IVon}
\end{equation}
In contrast to the off-resonance situation, $I$ does not follow a power law
even in the left-right symmetric model (see Fig.~\ref{fig:fig1}). 
Only for very large $V$ (or for $c\gg 1$), the second
term in the denominator of  (\ref{eq:IVon}) can be neglected and 
$I\propto V^{-2U_R}$
\cite{endnote2}. Thus, the voltage $V$ cannot be interpreted as a simple
infrared cutoff both for $\epsilon=\pm V/2$ and $U_L\neq U_R$ and the physics in
non-equilibrium is far more complex than in the linear-response limit \cite{expvalues}.

The analytic results (\ref{eq:flow}), (\ref{eq:IVoff}), and (\ref{eq:IVon}) 
derived from approximate FRG and RTRG-FS flow equations are confirmed by 
solving the full RG equations numerically.  The current for the left-right symmetric model
exhibits a power-law decay $I\propto V^{-2U}$ and thus a constant logarithmic
derivative only in the off-resonance case (see Fig.~\ref{fig:fig1}). 
Fig.~\ref{fig:fig2}(b) illustrates for $\epsilon=0$ and different
coupling asymmetries that the current from the full RTRG-FS flow
equation is captured by the analytic solution for the rates (\ref{eq:flow}) 
inserted in (\ref{eq:curr}). Moreover, the FRG compares nicely  
with accurate tDMRG  reference results obtained for large hoppings 
(see Fig.~\ref{fig:fig2}(a)).
Another transport property of experimental interest is the conductance
$G$, which as a function of the gate voltage $\epsilon$ most
importantly features the mentioned resonance at $\epsilon=\pm V/2$ as the
voltage becomes large (see Fig.~\ref{fig:fig3}).  As before, both RG
frameworks give agreeing numerical results for arbitrary values of $V/T_K$
and $\epsilon/T_K$, thus altogether providing reliable tools to study quantum
dot systems out of equilibrium.

\section{Time evolution}
\begin{figure}[t]
  \centering
  \includegraphics[width=0.9\linewidth,clip]{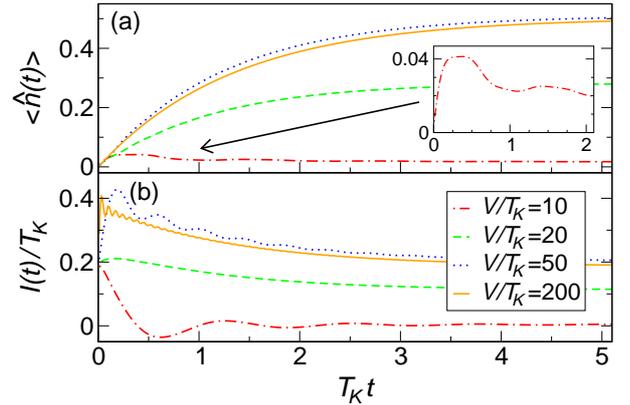}
  \caption{(Color online) Time evolution of the dot occupation $\left< \hat n(t)\right>$ and 
    the current $I(t)$ for $U_L=U_R=0.1/\pi$, $\epsilon=10\,T_K$, and the
    initial condition $\langle \hat{n}(0)\rangle=0$ \cite{zerocurrent}. At times $t\le 2/T_K$ we
    observe oscillating behavior.}
  \label{fig:fig4}
\end{figure}

The RTRG-FS allows for studying the time evolution
towards the steady state.  To this end, we initially prepare the system in a
state described by $\hat{\rho}(t<0)=\hat{\rho}_D^{(0)}\hat{\rho}_L\hat{\rho}_R$,
where $\hat{\rho}_D^{(0)}$ is an arbitrary initial density matrix of the dot and
$\hat{\rho}_{L/R}$ are grandcanonical distributions of the leads. At time
$t=0$, the coupling $H_{c}$ is suddenly switched on and transient dynamics
of $\hat \rho_D$ sets in. The latter can be fully described in terms of
$\Gamma_\alpha$ as a function of a Laplace variable $z$ which has to be
incorporated~\cite{RTlang}. By analytically
solving an approximation to these RG equations, one can obtain closed
integral representations both for the dot occupation $\langle \hat
n(t)\rangle$ and the current $I(t)$ by virtue of inverse Laplace
transform~\cite{PSS}. Numerical results for the time evolution are shown in
Fig.~\ref{fig:fig4}. We restrict ourselves to the left-right symmetric model for
simplicity. The long-time behavior away from resonance (i.e., at
$\epsilon,V,|\epsilon-V/2|\gg T_K, 1/t$) is given by
\begin{equation}
  \begin{split}
    \langle \hat n(t)\rangle &\approx \bigl(1-e^{-\Gamma_1t}\bigr)\langle\hat
    n\rangle-\frac{1}{2\pi}\,e^{-\Gamma_2 t}\,(T_Kt)^{1+2U}\\
    &\hspace{-8mm}\times
    \left[\frac{\sin\left[(\epsilon\!+\!\tfrac{V}{2})t\right]}
      {(\epsilon\!+\!\tfrac{V}{2})^2\,t^2}-
      \frac{\pi U}{4}\frac{\cos\left[(\epsilon\!+\!\tfrac{V}{2})t\right]}
      {(\epsilon\!+\!\tfrac{V}{2})^2\,t^2} + (V\!\to\!-V)\right]\!,
  \end{split}
\end{equation}

\noindent where $\langle\hat n\rangle$ follows from \eqref{eq:n}, and
$\Gamma_1\approx\Gamma$ as well as $\Gamma_2\approx \Gamma_1/2$ are two
decay rates.  Whereas $\Gamma_1$ describes the charge relaxation
process, $\Gamma_2$ is the broadening of the local level $\epsilon$ induced by
the coupling to the leads, i.e. it describes the relaxation of nondiagonal
elements of the local density matrix with respect to the charge states.  
We note that the dephasing rate $\Gamma_\phi=\Gamma_2-\Gamma_1/2\sim {\mathcal O}(U)$ 
is due to pure potential fluctuations on the dot and increases for
large Coulomb interactions. 
Most notable characteristics of the time evolution of 
both $\langle\hat n(t)\rangle$ as well as the current $I(t)$ are that (i) the relaxation
towards the stationary value is governed by both decay rates, (ii) the voltage
appears as an important energy scale for the dynamics setting the frequency of
an oscillatory behavior, and (iii) the exponential decay is accompanied by an
algebraic decay $\propto t^{2U-1}$.  The last result is of particular
importance for applications in error correction schemes of quantum information
processing as it contrasts the standard assumption of a purely exponential
decay~\cite{ECS}. We also note that in the short-time dynamics a reversal
of the current can occur (see the dashed-dotted curve in Fig.~\ref{fig:fig4}(b)). 
This effect is due to very 
strong charge fluctuations in the transient state, thus being impossible in 
systems with spin or orbital fluctuations~\cite{PSS}. Another interesting 
observation is that in the resonance case (dashed line) current oscillations 
are fully damped.

\section{Conclusion}We have studied non-equilibrium transport properties of a spinless 
single-level quantum dot coupled to leads via tunneling and Coulomb interaction, 
representing a fundamental model to describe the effects of charge fluctuations. 
Using two different RG methods we have presented analytic results
in the entire parameter regime and concluded that the
steady-state current $I(V)$ exhibits a power law only in specific cases. 
The one of highest experimental relevance is the situation of strong 
asymmetries in the tunneling couplings, where we generically observed a power law 
for large bias voltages $V$. Furthermore, the time evolution towards the steady state was studied. 
We found exponential decay on two different scales accompanied by voltage-dependent 
oscillations and power laws with interaction-dependent exponents.

We thank P.~Schmitteckert for providing the DMRG data of
Ref.~\cite{BSS}, and N.~Andrei, B.~Doyon, A.~Tsvelik, and A.~Zawadowski
for discussions.  This work was supported by the DFG-FG 723 and 912,
and by the AHV.



\begin{thebibliography}{99}
  \vspace*{-0.5cm}
\bibitem{S}
Schoeller, H., \textit{Eur. Phys. J. Special Topics}, {\bf 168}, 179 (2009).

\bibitem{B}
Mehta, P. \and Andrei, N., \textit{Phys.~Rev.~Lett.}, \textbf{96}, 216802 (2006); Erratum cond-mat/0703246.

\bibitem{BSZ} 
Borda, L., Vlad\'{a}r, K. \and Zawadowski, A., \textit{Phys. Rev. B}, \textbf{75}, 125107
(2007).

\bibitem{D}
Doyon, B., \textit{Phys.~Rev.~Lett.}, {\bf 99}, 076806 (2007).

\bibitem{BSS} Boulat, E., Saleur, H., \and Schmitteckert, P. \textit{Phys.~Rev.~Lett.},
  {\bf 101}, 140601 (2008).

\bibitem{BS}
Boulat, E. \and Saleur, H., \textit{Phys.~Rev.~B}, \textbf{77}, 033409 (2008). 

\bibitem{TD-NRG}
Anders, F. \and Schiller, A., \textit{Phys.~Rev.~Lett.}, {\bf 95},  196801  (2005).

\bibitem{TD-DMRG}
Daley, A.~{\it et al.}, \textit{J. Stat. Mech.}, 
P04005 (2004); White, S. \and Feiguin, A., \textit{Phys.~Rev.~Lett.}, {\bf 93}, 076401
(2004); Schmitteckert, P., \textit{Phys.~Rev.~B}, {\bf 70}, 121302 (2004);
Heidrich-Meisner, F., Feiguin, A., \and Dagotto, E., \textit{Phys.~Rev.~B}, {\bf 79},
235336 (2009).

\bibitem{W}
Weiss, S.~{\it et al.}, \textit{Phys.~Rev.~B}, {\bf 77}, 195316 (2008).

\bibitem{Sch}
Schmidt, T.~{\it et al.}, \textit{Phys.~Rev.~B}, {\bf 78}, 235110 (2008).

\bibitem{E}
Lesage, F. \and Saleur, H., \textit{Phys.~Rev.~Lett.}, {\bf 80}, 4370 (1998);
Schiller, A. \and Hershfield, S., \textit{Phys.~Rev.~B}, {\bf 62}, R16271 (2000);
Komnik, A., \textit{Phys.~Rev.~B}, {\bf 79}, 245102 (2009).

\bibitem{KS}
Keil, M. \and Schoeller, H., \textit{Phys.~Rev.~B}, \textbf{63}, 180302(R) (2001).

\bibitem{PSS}
Pletyukhov, M., Schuricht, D., \and Schoeller, H., \textit{Phys.~Rev.~Lett.}, {\bf 104}, 106801 (2010).

\bibitem{KT}
Lobaskin, D. \and Kehrein, S., \textit{Phys.~Rev.~B}, {\bf 71}, 193303 (2005); Hackl, A.~{\it et al.}, \textit{Phys.~Rev.~Lett.}, {\bf 102}, 196601 (2009);
Hackl, A., Vojta, M., \and Kehrein, S., \textit{Phys.~Rev.~B}, \textbf{80}, 195117 (2009).

\bibitem{FRGnonequi} Gezzi, R., Pruschke, Th., \and Meden, V., \textit{Phys.~Rev.~B},
  {\bf 75}, 045324 (2007); Jakobs, S., Meden, V., \and Schoeller, H.,
  \textit{Phys.~Rev.~Lett.}, {\bf 99}, 150603 (2007); Schmidt, H. \and W\"olfle, P.,
  \textit{Ann.~Phys.}~{\bf 19}, 60 (2010); Jakobs, S., Pletyukhov, M., \and Schoeller, H., \textit{Phys.~Rev.~B}, \textbf{81}, 195109 (2010).

\bibitem{RTRGKondo}
Schoeller, H. \and Reininghaus, F., \textit{Phys.~Rev.~B}, {\bf 80}, 045117 (2009);
Schuricht, D. \and Schoeller, H., \textit{Phys.~Rev.~B}, {\bf 80}, 075120 (2009).

\bibitem{KPBM}
Karrasch, C., Pletyukhov, M., Borda, L., and Meden, V., \textit{Phys.~Rev.~B}, {\bf 81}, 125122 (2010).

\bibitem{RTlang} 
Andergassen, S.~\emph{et al.}, (in preparation).

\bibitem{chinrg} The power-law scaling of $\chi_{V=0}^\textnormal{sym}$ 
  as a function of $\Gamma_L^0=\Gamma_R^0$ \cite{BSZ} is
  captured both by FRG 
  and RTRG-FS in good agreement with NRG data.
  
\bibitem{endnote2} In contrast, the perturbative study of
  Ref.~\cite{D} yields $I\propto V^{-U}$, with $U=U_L=U_R$.

\bibitem{expvalues}
In quantum dot transport experiments, the bandwidth $B\sim 1eV$ is typically large compared
to the Kondo scale $T_K\sim 0.2\textnormal{meV}$ which is in turn larger than the usual environment temperature $T\sim 0.02\textnormal{meV}$. Thus, the regime of negative differential conductance of the IRLM is roughly associated with a current of the order of $I\sim 0.5\textnormal{nA}$ and voltages $V\sim 20\textnormal{meV}$.

\bibitem{zerocurrent}
The current in the lead at times  $t\sim 1/B$ is determined 
by the nonzero displacement current $d \langle\hat n(t)\rangle/dt$.

\bibitem{ECS}
Peskill, J.~ in \emph{Introduction to Quantum Computation and Information}
(H.-K. Lo, S. Popescu, and T. Spiller, World Scientific, Singapore, 1998);
Fischer, J.~ \and Loss, D., \textit{Science}, \textbf{324}, 1277 (2009).


\end{thebibliography}
\end{document}